*Working Paper*

# Prediction of retail chain failure: examples of recent U.S. retail failures

Shawn Berry, DBA[1*]

March 13, 2024

[1]William Howard Taft University, Lakewood CO, USA

*Correspondence: shawn.berry.8826@taftu.edu

**Abstract**  Over the last several years, several well-established and prominent brick-and-mortar retail chains have ceased operations, raising concerns for something that some have referred to as a retail apocalypse. While the demise of brick-and-mortar is far from certain, scholars have attempted to model the likelihood that a retailer is about to fail using different approaches. This paper examines the failures of Bed Bath and Beyond, J.C. Penney, Rite Aid, and Sears Holdings in the United States between 2013 and 2022. A model of retail failure is presented that considers internal and external firm factors using both annual report and macroeconomic data. The findings suggest that certain revenue-based financial ratios and the annual average U.S. inflation rates are statistically significant predictors of failure. Furthermore, the failure model demonstrated that it can provide a nontrivial early warning signal at least the year before failure. The paper concludes with a discussion and directions for future research.

---

## 1. Introduction

The number of high profile retail chain failures over the last few years, including prior to the COVID-19 pandemic, have created concerns for not just employees of these retailers but also suppliers, investors, and property managers that become exposed to the financial consequences of such failures. Scholars have often referred to this failure phenomenon as a retail apocalypse (Peterson, 2017; Mende, 2019; Kaulfinger and Neuenschwander, 2020). The sources of retail chain failure are various. In fact, Habib et al. (2020) identify three broad categories of factors, specifically " (i) firm-level fundamental determinants, (ii) macroeconomic determinants and (iii) firm-level corporate governance determinants" (p.1023). However, since the underlying reasons for failure are unique to the circumstances of any one retailer and seldom the same for other firms, some studies have pointed to aspects of a business that may precipitate failure. Habib et al. (2020) identify five such categories of circumstances, namely, " (i) financial reporting and auditing consequences, (ii) firm-level operational consequences, (iii) capital market consequences and (iv) corporate governance consequences" (p.1023). For example, Kaufinger and Neuenschwander (2020) suggest that a retailer may be able to avoid failure through the consideration of which accounting methodology is used for inventory valuation. Redd and Vickerie (2017) suggest that technological change for traditional brick-and-mortar retailers to transition to e-commerce was perhaps a challenge. While not a singular cause for retail failure, such studies add valuable insights into the knowledge of the general reasons for retail failure. Therefore, the ability to anticipate and predict the possibility of failure is important to those that are ultimately affected to mitigate risk (Keener, 2013).

As a result of these collected insights, prediction models have been created by scholars to quantify the likelihood of retail firm failure. Some models rely on ratios based on assets, such as the Altman model (Altman et al., 2014), and other models generally use a variety and combination of other firm financial ratios (Brédart, 2014; Evans and Mathur, 2014). Furthermore, other models incorporate internal firm cost ratios and macroeconomic indicators as external factors for failure (Ceylan, 2021; Keener, 2013).

In this short paper, data from recent high profile, multi-store retail chain failures between 2013 and 2022, specifically, Bed Bath and Beyond, J.C. Penney, Rite Aid and Sears Holdings, are used to create a model of retail failure. Such a model is valuable to business leaders and analysts to help assess the possibility of bankruptcy risk, and assist in the mitigation of potential loss and strains of business relationships connected to a failing firm. Selected key financial data are drawn from annual reports to construct internal firm factors and selected macroeconomic indicators are used as external firm factors. The correlates of failure are explored, and finally, a predictive model of failure is created using significant internal and external firm factors to estimate probabilities of failure for each of the firms over the sample period. Unlike other studies that incorporate paired samples of failed and non-failed forms to construct their models (Altman et al., 2014; Brédart, 2014; Hayes, Hodge and Hughes, 2010; Mcgurr and Devaney, 1998), the sample data form each firm is coded for each year prior to their final year when failure occurs as having not failed. Finally, the study will conclude with a discussion of the results and directions for future research,

## 2. Materials and Methods

The sample data used in this study has been drawn from various secondary data sources. First, company data was drawn from publicly available SEC 10-K annual report filings for Bed Bath and Beyond (2015-2022), J.C. Penney (2013-2020), Rite Aid (2013-2022) and Sears Holdings (2013-2018) for revenue, costs, and store counts to examine internal factors of failure. Data for external or macroeconomic factors, specifically, the annual average US interest rate and inflation rates, and consumer satisfaction ratings for each of the chains as published by the American Consumer Satisfaction Index (ACSI). Failure of a firm was coded as 0 in years when that firm was still in operation, and coded as 1 for the year that failure occurred. Costs were then calculated as ratios of revenue for each year. Revenue was used as a measure of sales performance. Cost of revenue and selling and general administration (SGA) costs were used as measures of how effective the firm is at controlling its own internal costs for selling. Long-term debt was used as a measure of the potential for debt default. Finally, EBITDA was used as a measure of the profitability of the firm.
.

## 3. Results

The sample data was analyzed and modelled.

### *3.1. Descriptive statistics*

Table 1 summarizes the descriptive statistics for the sample data with the means, standard deviations and Shapiro-Wilk test statistics. Revenue appears to have the highest variation for retail firms. Among cost factors, the cost of revenue has the highest variation, followed by SGA and long-term debt.

**Table 1.** Descriptive statistics

| Variable | Mean | Standard deviation | Shapiro-Wilk (p-value) |
|---|---|---|---|
| Revenue | 16854.81 (M$) | 8105.03 | 0.945 (0.107) |
| SGA | 4450.66 (M$) | 1730.76 | 0.929 (0.036) |
| Cost of revenue | 12301.91 (M$) | 6545.78 | 0.909 (0.011) |
| EBITDA | 191.16 (M$) | 955.70 | 0.934 (0.050) |
| Stores | 1891.81 | 1194.25 | 0.816 (<0.001) |
| US interest rate | 2.16% | 0.58 | 0.921 (0.022) |
| US inflation rate | 2.11% | 1.85 | 0.720 (<0.001) |
| ACSI score | 76.31 | 3.11 | 0.959 (0.257) |
| Long-term debt | 3475.12 (M$) | 1680.49 | 0.934 (0.049) |
| Pandemic | 0.22 | 0.42 | 0.512 (<0.001) |
| SGA/Revenue | 0.29 | 0.07 | 0.938 (0.066) |
| Cost of revenue/Revenue | 0.71 | 0.7 | 0.915 (0.016) |
| EBITDA/Revenue | -0.012 | 0.14 | 0.625 (<0.001) |
| Long-term debt/Revenue | 0.30 | 0.50 | 0.339 (<0.001) |

**Source:** data analysis

### 3.2 *Sample data*

The sample data is presented in Table 2 below.

**Table 2.** Retail chain failures and key firm financial and macroeconomic indicators

| Chain | Year | Fail 1=yes | Revenue ($M) | Cost of revenue ($M) | SGA ($M) | EBITDA ($M) | Stores | US interest rate (%) | US inflation rate (%) | SGA/ Revenue | Cost of revenue/ Revenue | Long-term debt ($M) | Long-term debt/ Revenue | Pandemic 1= yes | ACSI score | EBITDA/ Revenue |
|---|---|---|---|---|---|---|---|---|---|---|---|---|---|---|---|---|
| Bed Bath & Beyond | 2015 | 0 | 12,104 | 7,484 | 3,205 | 1,689 | 1513 | 2.1 | 0.12 | 0.26 | 0.62 | 1500 | 0.12 | 0 | 75 | 0.14 |
| Bed Bath & Beyond | 2016 | 0 | 12,216 | 7,639 | 3,441 | 1,426 | 1530 | 1.8 | 1.26 | 0.28 | 0.63 | 1492 | 0.12 | 0 | 79 | 0.12 |
| Bed Bath & Beyond | 2017 | 0 | 12,349 | 7,906 | 3,682 | 1,074 | 1546 | 2.3 | 2.13 | 0.3 | 0.64 | 1492 | 0.12 | 0 | 76 | 0.09 |
| Bed Bath & Beyond | 2018 | 0 | 12,029 | 7,925 | 3,681 | 251.69 | 1552 | 2.9 | 2.44 | 0.31 | 0.66 | 1488 | 0.12 | 0 | 79 | 0.02 |
| Bed Bath & Beyond | 2019 | 0 | 11,159 | 7,617 | 3,732 | -357.55 | 1533 | 2.1 | 1.81 | 0.33 | 0.68 | 1488 | 0.13 | 0 | 80 | -0.03 |
| Bed Bath & Beyond | 2020 | 0 | 9,233 | 6,115 | 3,224 | 81.06 | 1500 | 0.9 | 1.23 | 0.35 | 0.66 | 1488 | 0.16 | 1 | 79 | 0.01 |
| Bed Bath & Beyond | 2021 | 0 | 7,868 | 5,384 | 2,692 | -114.33 | 1020 | 1.5 | 4.7 | 0.34 | 0.68 | 1190 | 0.15 | 1 | 80 | -0.01 |
| Bed Bath & Beyond | 2022 | 1 | 5,345 | 4,130 | 2,373 | -2,992.29 | 953 | 3 | 8 | 0.44 | 0.77 | 1180 | 0.22 | 1 | 78 | -0.56 |
| Rite Aid | 2013 | 0 | 25,526 | 18,203 | 6,561 | 1,079 | 4570 | 2.4 | 1.62 | 0.26 | 0.71 | 5708 | 0.22 | 0 | 74 | 0.04 |
| Rite Aid | 2014 | 0 | 26,528 | 18,952 | 6,696 | 1,241 | 4623 | 2.5 | 1.46 | 0.25 | 0.71 | 5459 | 0.21 | 0 | 78 | 0.05 |
| Rite Aid | 2015 | 0 | 20,770 | 15,778 | 4,581 | 762.24 | 4561 | 2.1 | 0.12 | 0.22 | 0.76 | 6967 | 0.34 | 0 | 69 | 0.04 |
| Rite Aid | 2016 | 0 | 22,928 | 17,863 | 4,777 | 655.92 | 4536 | 1.8 | 1.26 | 0.21 | 0.78 | 3273 | 0.14 | 0 | 78 | 0.03 |
| Rite Aid | 2017 | 0 | 21,529 | 16,749 | 4,651 | 1,838 | 2550 | 2.3 | 2.13 | 0.22 | 0.78 | 3371 | 0.16 | 0 | 77 | 0.09 |
| Rite Aid | 2018 | 0 | 21,640 | 16,963 | 4,592 | 240.87 | 2469 | 2.9 | 2.44 | 0.21 | 0.78 | 3479 | 0.16 | 0 | 76 | 0.01 |
| Rite Aid | 2019 | 0 | 21,928 | 17,202 | 4,587 | 493.37 | 2461 | 2.1 | 1.81 | 0.21 | 0.78 | 5807 | 0.26 | 0 | 75 | 0.02 |
| Rite Aid | 2020 | 0 | 24,043 | 19,339 | 4,657 | 417.45 | 2510 | 0.9 | 1.23 | 0.19 | 0.8 | 5909 | 0.25 | 1 | 72 | 0.02 |
| Rite Aid | 2021 | 0 | 24,568 | 19,462 | 5,034 | -54.97 | 2450 | 1.5 | 4.7 | 0.2 | 0.79 | 5345 | 0.22 | 1 | 71 | 0.00 |
| Rite Aid | 2022 | 1 | 24,092 | 19,288 | 4,902 | -255.42 | 2450 | 3 | 8 | 0.2 | 0.8 | 5311 | 0.22 | 1 | 80 | -0.01 |
| Sears Holdings | 2013 | 0 | 36188 | 27433 | 9384 | -487 | 2429 | 2.4 | 1.62 | 0.26 | 0.76 | 2531 | 0.07 | 0 | 77 | -0.01 |
| Sears Holdings | 2014 | 0 | 31198 | 24049 | 8220 | -718 | 1725 | 2.5 | 1.46 | 0.26 | 0.77 | 2878 | 0.09 | 0 | 73 | -0.02 |
| Sears Holdings | 2015 | 0 | 25146 | 19336 | 6857 | -836 | 1672 | 2.1 | 0.12 | 0.27 | 0.77 | 1971 | 0.08 | 0 | 71 | -0.03 |
| Sears Holdings | 2016 | 0 | 22138 | 15184 | 6109 | -808 | 1430 | 1.8 | 1.26 | 0.28 | 0.69 | 3470 | 0.16 | 0 | 77 | -0.04 |

| | | | | | | | | | | | | | | | |
|---|---|---|---|---|---|---|---|---|---|---|---|---|---|---|---|
| Sears Holdings | 2017 | 0 | 16702 | 11349 | 5139 | -562 | 1002 | 2.3 | 2.13 | 0.31 | 0.68 | 2199 | 0.13 | 0 | 73 | -0.03 |
| Sears Holdings | 2018 | 1 | 6709 | 5899 | 2626 | -571 | 332 | 2.9 | 2.44 | 0.39 | 0.88 | 2239 | 0.33 | 0 | 73 | -0.09 |
| JC Penney | 2013 | 0 | 11859 | 8367 | 4114 | -819 | 1094 | 2.4 | 1.62 | 0.35 | 0.71 | 4839 | 0.41 | 0 | 79 | -0.07 |
| JC Penney | 2014 | 0 | 12257 | 7996 | 3993 | 323 | 1062 | 2.5 | 1.46 | 0.33 | 0.65 | 5227 | 0.43 | 0 | 77 | 0.03 |
| JC Penney | 2015 | 0 | 12625 | 8074 | 3775 | 654 | 1021 | 2.1 | 0.12 | 0.3 | 0.64 | 4668 | 0.37 | 0 | 74 | 0.05 |
| JC Penney | 2016 | 0 | 12547 | 8071 | 3538 | 926 | 1013 | 1.8 | 1.26 | 0.28 | 0.64 | 4339 | 0.35 | 0 | 82 | 0.07 |
| JC Penney | 2017 | 0 | 12554 | 8208 | 3845 | 935 | 872 | 2.3 | 2.13 | 0.31 | 0.65 | 3780 | 0.30 | 0 | 79 | 0.07 |
| JC Penney | 2018 | 0 | 11664 | 7870 | 3596 | 568 | 864 | 2.9 | 2.44 | 0.31 | 0.67 | 3716 | 0.32 | 0 | 77 | 0.05 |
| JC Penney | 2019 | 0 | 10716 | 7013 | 3585 | 583 | 849 | 2.1 | 1.81 | 0.33 | 0.65 | 3826 | 0.36 | 0 | 78 | 0.05 |
| JC Penney | 2020 | 1 | 1196 | 813 | 572 | -546 | 846 | 0.9 | 1.23 | 0.48 | 0.68 | 3574 | 2.99 | 1 | 76 | -0.46 |

### *3.3. Regression analysis:*

The sample data was analyzed using logistic regression analysis..

Figure 1 illustrates the correlation matrix for the dataset. The dependent variable fail appears to be strongly correlated with EBITDA/revenue, long-term debt/revenue, the annual average US inflation rate, if the year was a COVID-19 pandemic year (yes=1), and SGA/revenue. However, among the strongly correlated independent variables, there appears to be some multicollinearity.

**Figure 1.** Correlation matrix of dataset

|  | Year | Fail (1=yes) | Revenue ($M) | Cost of revenue ($M) | SGA ($M) | EBITDA ($M) | Number of stores | US interest rate (%) | US inflation rate (%) | SGA/ Revenue | Cost of revenue/ Revenue | Long-term debt ($M) | EBITDA/ Revenue | Long-term debt/ Revenue | Pandemic (1= yes) | ACSI score |
|---|---|---|---|---|---|---|---|---|---|---|---|---|---|---|---|---|
| Year | 1.00 | 0.49 | -0.39 | -0.32 | -0.55 | -0.31 | -0.28 | -0.20 | 0.67 | 0.19 | 0.20 | -0.18 | -0.40 | 0.18 | 0.75 | 0.17 |
| Fail (1=yes) | 0.49 | 1.00 | -0.36 | -0.28 | -0.41 | -0.52 | -0.24 | 0.19 | 0.58 | 0.50 | 0.39 | -0.09 | -0.73 | 0.49 | 0.49 | 0.05 |
| Revenue ($M) | -0.39 | -0.36 | 1.00 | 0.99 | 0.93 | 0.11 | 0.65 | 0.12 | -0.11 | -0.79 | 0.46 | 0.38 | 0.35 | -0.42 | -0.21 | -0.35 |
| Cost of revenue ($M) | -0.32 | -0.28 | 0.99 | 1.00 | 0.90 | 0.06 | 0.65 | 0.11 | -0.05 | -0.78 | 0.56 | 0.39 | 0.29 | -0.38 | -0.14 | -0.38 |
| SGA ($M) | -0.55 | -0.41 | 0.93 | 0.90 | 1.00 | -0.01 | 0.48 | 0.21 | -0.17 | -0.58 | 0.31 | 0.21 | 0.32 | -0.47 | -0.34 | -0.28 |
| EBITDA ($M) | -0.31 | -0.52 | 0.11 | 0.06 | -0.01 | 1.00 | 0.35 | -0.14 | -0.48 | -0.49 | -0.35 | 0.26 | 0.79 | -0.12 | -0.39 | 0.04 |
| Number of stores | -0.28 | -0.24 | 0.65 | 0.65 | 0.48 | 0.35 | 1.00 | -0.01 | -0.12 | -0.65 | 0.34 | 0.50 | 0.25 | -0.19 | -0.10 | -0.26 |
| US interest rate (%) | -0.20 | 0.19 | 0.12 | 0.11 | 0.21 | -0.14 | -0.01 | 1.00 | 0.35 | -0.03 | 0.19 | -0.05 | 0.01 | -0.37 | -0.45 | 0.08 |
| US inflation rate (%) | 0.67 | 0.58 | -0.11 | -0.05 | -0.17 | -0.48 | -0.12 | 0.35 | 1.00 | 0.12 | 0.32 | -0.10 | -0.45 | -0.09 | 0.59 | 0.29 |
| SGA/ Revenue | 0.19 | 0.50 | -0.79 | -0.78 | -0.58 | -0.49 | -0.65 | -0.03 | 0.12 | 1.00 | -0.32 | -0.46 | -0.67 | 0.54 | 0.20 | 0.27 |
| Cost of revenue/ Revenue | 0.20 | 0.39 | 0.46 | 0.56 | 0.31 | -0.35 | 0.34 | 0.19 | 0.32 | -0.32 | 1.00 | 0.25 | -0.24 | -0.11 | 0.20 | -0.43 |
| Long-term debt ($M) | -0.18 | -0.09 | 0.38 | 0.39 | 0.21 | 0.26 | 0.50 | -0.05 | -0.10 | -0.46 | 0.25 | 1.00 | 0.18 | 0.13 | -0.02 | -0.31 |
| EBITDA/ Revenue | -0.40 | -0.73 | 0.35 | 0.29 | 0.32 | 0.79 | 0.25 | 0.01 | -0.45 | -0.67 | -0.24 | 0.18 | 1.00 | -0.57 | -0.51 | 0.00 |
| Long-term debt/ Revenue | 0.18 | 0.49 | -0.42 | -0.38 | -0.47 | -0.12 | -0.19 | -0.37 | -0.09 | 0.54 | -0.11 | 0.13 | -0.57 | 1.00 | 0.32 | -0.01 |
| Pandemic (1= yes) | 0.75 | 0.49 | -0.21 | -0.14 | -0.34 | -0.39 | -0.10 | -0.45 | 0.59 | 0.20 | 0.20 | -0.02 | -0.51 | 0.32 | 1.00 | 0.04 |
| ACSI score | 0.17 | 0.05 | -0.35 | -0.38 | -0.28 | 0.04 | -0.26 | 0.08 | 0.29 | 0.27 | -0.43 | -0.31 | 0.00 | -0.01 | 0.04 | 1.00 |

**Source:** data analysis

The impact of selected factors that are external to the firm were examined with respect to the likelihood of retail chain failure. The dependent variable, fail, was once again modelled using logistic regression individually. The American Consumer Satisfaction Index (ACSI) score for the retail chains represents a proxy for consumer preference for the retail chains. The existence of the COVID-19 pandemic, a binary dummy variable (1= yes), was examined and non-COVID-19 pandemic years were coded with zero. Finally, the annual average US interest rate and the annual average US inflation rates were used to evaluate the effects of the macroeconomy on the likelihood of retail chain failure.

Table 3 below displays the results of the analysis. Surprisingly, ACSI was not a statistically significant predictor for failure. The most statistically significant model had the annual average US inflation rate as an independent variable, with both the intercept and slope parameters being significant, $p<0.001$ and $p=0.05$ levels, respectively. The model that evaluates the effect of the existence of a pandemic in any given year on failure also had a statistically significant intercept and slope parameter at $p=0.001$ and $p=0.05$ levels, respectively. Finally, the model that evaluates the effect of the annual average US interest rates had a statistically significant intercept coefficient at the $p=0.1$ level.

The pandemic model implies that when there is a pandemic, retail chains are over 17 times more likely to fail than when there is not a pandemic. The inflation model implies that with a 1% increase in the annual average US rate of inflation, retail chains are twice as likely to fail. Finally, the interest rate model implies that with a 1% increase in the annual average US interest rates, retail firms are just over 3.5 times more likely to fail. The ACSI model implies that changes in the ACSI score for a retail chain do not significantly change the likelihood of failure for the firm (exp(0.0555) = 1.05), and that consumer attitudes toward a chain is not a singular determinant of the likelihood of failure.

**Table 3.** Logistic regression results with external factors. Dependent variable: fail.

| Estimates | ACSI | Pandemic | US interest rate | US inflation rate |
|---|---|---|---|---|
| Intercept | -6.191 | -3.178 (**) | -4.856 (.) | -3.957 (***) |
| (p-value) | (0.657) | (0.0019) | (0.096) | (<0.001) |
| [s.e.] | [13.947] | [1.021] | [2.918] | [1.164] |
| | | | | |
| Slope | 0.0555 | 2.890 (*) | 1.267 | 0.708 (*) |
| (p-value) | (0.760) | (0.023) | (0.286) | (0.021) |
| [s.e.] | [0.182] | [1.275] | [1.187] | [0.307] |
| AIC | 28.017 | 21.958 | 26.757 | 20.349 |

Signif. codes :  0 '***' 0.001 '**' 0.01 '*' 0.05 '.' 0.1 ' ' 1

**Source :** data analysis

The effect of selected internal factors for a retail chain on the likelihood of failure, represented by the dependent variable fail, were examined using logistic regression modelling. Specifically, revenue and various measures of cost were used as independent variables. The cost of revenue and selling and general administration (SGA) were used as proxies to represent the ability to control costs by the firm. The number of stores represents the scale of operations of the retail chain. Finally, long-term debt was used as a measure of the level of indebtedness of the firm beyond the costs of selling and generating revenue.

The results are illustrated in Table 4. The failure model with EBITDA as an independent variable was the only model with the lowest AIC, and where the intercept ($p=0.004$) and slope ($p=0.086$) parameters were both statistically significant at 0.001 and 0.1, respectively. Although the intercepts are not statistically significant, Revenue and SGA are statistically significant at 0.1 ($p=0.079$ and $p=0.057$, respectively). The number of stores did not suggest that the physical scale of the retail chain operations was significant on its own. More surprisingly, the level of long-term debt and levels of cost of revenue also were not statistically significant predictors of failure.

**Table 4.** Logistic regression results with internal factors. Dependent variable: fail.

| Estimates | Revenue | SGA | Cost of revenue | Stores | EBITDA | Long-term debt |
|---|---|---|---|---|---|---|
| Intercept | 0.757 | 3.546 | -0.114 | -0.02467 | -2.494 (**) | -1.374 |
| (p-value) | (0.583) | (0.193) | (0.009) | (0.986) | (0.0036) | (0.247) |
| [s.e.] | [1.379] | [2.725] | [3.576] | [1.376] | [0.858] | [1.186] |
| Slope | -0.0002 (.) | -0..0015 (.) | -0.00019 | -0.0013 | -0.0022 (.) | -0.0002 |
| (p-value) | (0.079) | (0.057) | (0.154) | (0.207) | (0.086) | (0.608) |
| [s.e.] | [0.0001] | [0.0008] | [0.0001] | [0.001] | [0.0013] | [0.0003] |
| AIC | 23.039 | 20.172 | 25.095 | 25.277 | 19.806 | 27.841 |

Signif. codes :  0 '***' 0.001 '**' 0.01 '*' 0.05 '.' 0.1 ' ' 1

**Source :** data analysis

    The logistic regression models of retail chain failure were once again recreated by transforming key independent variables to fractions of revenue. Table 5 below illustrates the results of the analysis. All models had intercepts and slope parameters that were statistically significant. The model with the highest levels of statistical significance had SGA/Revenue as an independent variable, with the intercept being significant at 0.001 (p=0.009) and the slope parameter significant at 0.05 (p=0.028). The model with EBITDA/Revenue, although having a lower AIC, similarly had an intercept that was significant at 0.001 and the slope was significant at 0.1. The failure model using SGA/Revenue implies that for a 1 unit increase in this ratio, the likelihood of failure will increase substantially.

**Table 5.** Logistic regression results with internal factors as ratios of revenue. Dependent variable: fail.

| Estimates | SGA/ Revenue | Cost of revenue/ Revenue | EBITDA/ Revenue | Long-term debt/ Revenue |
|---|---|---|---|---|
| Intercept | -9.245 (**) | -17.021 (*) | -3.264 (**) | -3.238 (*) |
| (p-value) | (0.009) | (0.043) | (0.009) | (0.041) |
| [s.e.] | [3.576] | [8.404] | [1.259] | [1.584] |
| Slope | 22.728 (*) | 20.253 (.) | -41.769 (.) | 4.274 |
| (p-value) | (0.028) | (0.064) | (0.089) | (0.459) |
| [s.e.] | [10.357] | [10.932] | [24.599] | [5.768] |
| AIC | 20.76 | 23.131 | 13.951 | 23.175 |

Signif. codes :  0 '***' 0.001 '**' 0.01 '*' 0.05 '.' 0.1 ' ' 1

**Source :** data analysis

    Using the insights from the previous analyses, a failure prediction model was created for the dependent variable fail. Unfortunately, since the use of binomial logistic regression was not effective in this case due to convergence problems, the model was estimated using penalized maximum likelihood logistic regression. The results of the analysis are displayed in Table 6. The independent variables used are the US

inflation rate, long-term debt/revenue, and EBITDA/revenue. The Wald test statistic for the model is statistically significant ($p<0.05$), implying that the variables are good contributors to the fit of the model.

**Table 6.** Failure prediction model. Dependent variable: fail.

|  | Estimate | Std. Error | Chi-sq. | Pr (>|t|) |
|---|---|---|---|---|
| Intercept | -4.349 | 1.434 | 9.207 | 0.002 |
| US inflation rate | 0.592 | 0.297 | 3.971 | 0.046 |
| Long-term debt/Revenue | 1.374 | 1.057 | 1.689 | 0.194 |
| EBITDA/Revenue | -1.606 | 4.303 | 0.139 | 0.709 |

Likelihood ratio test=13.81581 on 3 df, p=0.003166889, n=32
Wald test = 12.7042 on 3 df, p = 0.005321982

Signif. codes :  0 '***' 0.001 '**' 0.01 '*' 0.05 '.' 0.1 ' ' 1

The failure prediction model in Table 6 was used to estimate the probability of failure for each retail chain in each year of operation using data for the independent variables from the dataset. The results of this estimation process is summarized by chain and year below in Table 7. The probabilities are calculated by taking the odds ratio of the result of the dependent variable fail, as given by

$$P(fail) = \exp(fail)/(1+\exp(fail)).$$

**Table 7.** Estimates of probability of failure for retail chains using failure prediction model

| Year | Bed Bath & Beyond | Rite Aid | Sears Holdings | J.C. Penney |
|---|---|---|---|---|
| 2013 | - | 0.063 | 0.042 | 0.130 |
| 2014 | - | 0.054 | 0.042 | 0.110 |
| 2015 | 0.017 | 0.039 | 0.019 | 0.043 |
| 2016 | 0.033 | 0.040 | 0.047 | 0.072 |
| 2017 | 0.056 | 0.063 | 0.070 | 0.101 |
| 2018 | 0.074 | 0.085 | 0.162 | 0.129 |
| 2019 | 0.059 | 0.082 | ∗ | 0.103 |
| 2020 | 0.043 | 0.056 | ∗ | 0.998 |
| 2021 | 0.263 | 0.304 | ∗ | ∗ |
| 2022 | 0.884 | 0.760 | ∗ | ∗ |

- : not available
∗: firm has ceased operations

## 4. Discussion

.
The probability estimates generated by the failure model has several implications. First, the failure model appears to signal in the year before failure that there is a non-trivial chance of the firm becoming bankrupt. This early warning signal is observed for Bed Bath & Beyond and Rite Aid. Second, the probability estimates for J.C. Penney suggest that the firm was able to drive the likelihood of failure down to just over 4% in 2015 and hovered variously about 10-13% chance in the years preceding its failure, implying that J.C. Penney was making efforts to be solvent but failed catastrophically when it didn't need to. Third, the failure model did not provide a dramatic early warning signal for Sears Holdings as the firm had essentially hovered around a 4% chance of failure in prior years, jumping only to a 7% chance a year before failure. The Sears result implies that their failure is a result of factors beyond the model, such as governance, for example.

The results in Table 1 generally suggest that retail firms are susceptible to enormous variation in revenue, and their profitability is also threatened by the large variation in the cost of revenue, SGA, and long-term debt. These factors together imply that the lack of control of the variance of cost factors and uncertainty of revenue are contributions to the failure of retailers.

The results in Table 5 suggest that financial ratios that use costs as a fraction of revenue are generally statistically significant indicators of internal factors that may predict to firm failure. The results in Table 3 suggest that the annual average U.S. inflation rate is a statistically significant external factor that may help to predict firm failure in addition to internal factors.

## 5. Directions for future research

The possibility of retail failure remains a fact of life for any retail firm. Additional research is required to enable the development of tools to provide early warning signals of possible failure, similar to the failure prediction model that is presented in this paper. Advance warning that a retail chain is possibly going to fail based on recent indicators is quite valuable to investors and suppliers. Therefore, more research is required to leverage publicly available data that implies the liquidity status of a company with respect to the potential for failure based on peers in the retail industry. This implies the importance of competitive benchmarking of performance as early indicators of success or failure provided that this information can be tied to a firm's financial ratios. Finally, future research must be done to identify sources of variance that contribute to poor control of SGA and cost of revenue as triggers that could precipitate a process of failure for a firm.

## 6. Limitations

As with any research, limitations must be acknowledged. First, since secondary data sources were used for firm financial statements, ACSI scores, and macroeconomic data, changes in reporting or computation used by those sources may create hidden biases. This also includes non-GAAP reporting of common financial information, resulting in potential inconsistencies. Second, the completeness of reporting when a firm ceases operations presented a limitation, particularly for J.C. Penney in their final year where a 10-Q was filed rather than a 10-K. Finally, since ACSI scores were not available for Bed Bath and Beyond before 2015, the financial data for that firm was collected from 2015 onward.

## 7. Patents

There are no patents resulting from the work reported in this manuscript.

## 8. Funding

This research received no external funding.

## 9. Conflicts of Interest

The authors declare no conflict of interest.